\documentclass[useAMS,usenatbib]{mn2e}
\usepackage{graphicx,amssymb, txfonts}
\usepackage[dvips]{epsfig}
\usepackage{amssymb}
\usepackage{lscape}
\usepackage{subfig}
\usepackage{float}
\usepackage{color}
\usepackage{graphicx}
\usepackage{verbatim}
\usepackage{rotating}
\usepackage{ulem}

\def\H2{H$_2$}
\def\p1{Paper~I}

\def\kms {$\rm km\,s^{-1}$}

\def\xy{x$_{\rm y}$}
\def\xyi{x$_{\rm yi}$}
\def\xio{x$_{\rm io}$}
\def\xo{x$_{\rm o}$}

\def\my{m$_{\rm y}$}
\def\myi{m$_{\rm yi}$}
\def\mio{m$_{\rm io}$}
\def\mo{m$_{\rm o}$}
\def\starlight{{\sc starlight}}

\title[2D stellar populations in Mrk573]{Disentangling the near infrared continuum spectral components of the inner 500 pc of Mrk 573: two-dimensional maps}

\author[Diniz et al.]{M. R. Diniz$^{1}$\thanks{E-mail:
diniz.mr@gmail.com}, R. A. Riffel$^{1}$, R. Riffel$^{2}$,  D. M. Crenshaw$^{4}$,  T. Storchi-Bergmann$^{2}$, \newauthor T. C. Fischer$^{3,4}$,  H. R. Schmitt$^{5}$, S. B. Kraemer$^{6}$ \\
$^{1}$ Universidade Federal de Santa Maria, Departamento de F\'\i sica, Centro de Ci\^encias Naturais e Exatas, 
97105-900, Santa Maria, RS, Brazil\\
$^{2}$ Universidade Federal do Rio Grande do Sul, Instituto de F\'\i sica, CP 15051, Porto Alegre 91501-970, RS, Brazil\\
$^{3}$ Astrophysics Science Division, Goddard Space Flight Center, Code 665, Greenbelt, MD 20771, USA \\
$^{4}$ Department of Physics and Astronomy, Georgia State University, Astronomy Offices, 25 Park Place, Suite 605, Atlanta, GA 30303, USA \\
$^{5}$ Naval Research Laboratory, Washington, DC 20375, USA 0000-0001-7376-8481\\
$^{6}$ Institute for Astrophysics and Computational Sciences, Department of Physics, The Catholic University of America, Washington,DC 20064, USA.
}
\begin{document}


\pagerange{\pageref{firstpage}--\pageref{lastpage}} \pubyear{2013}

\maketitle

\label{firstpage}

\begin{abstract}

We present a near infrared study of the spectral components of the continuum in the inner 500$\times$500~pc$^2$ of the nearby Seyfert galaxy Mrk\,573 using adaptive optics near-infrared integral field spectroscopy with the instrument NIFS of the Gemini North Telescope at a spatial resolution of $\sim$50~pc. We performed spectral synthesis using the {\sc starlight} code and constructed maps for the contributions of different age components of the stellar population: young ($age\leq100$~Myr), young-intermediate ($100<age\leq700$~Myr), intermediate-old ($700$~Myr $<age\leq2$~Gyr) and old ($age>2$~Gyr) to the near-IR K-band continuum, as well as their contribution to the total stellar mass. We found that the old stellar population is dominant within the inner 250~pc, while the intermediate age components dominate the continuum at larger distances. A young stellar component contributes up to $\sim$20\% within the inner $\sim$\,70\,pc, while hot dust emission and  featureless continuum components are also necessary to fit the nuclear spectrum, contributing up to 20\% of the K-band flux there. The radial distribution of the different age components in the inner kiloparsec of Mrk\,573 is similar to those obtained by our group for the Seyfert galaxies Mrk\,1066, Mrk\,1157 and NGC\,1068 in previous works using a similar methodology.  Young stellar populations ($\leq$100~Myr) are seen in the inner 200--300~pc for all galaxies contributing with $\ge$\,20\% of the K-band flux, while the near-IR continuum is dominated by the contribution of intermediate-age stars ($t=$100~Myr--2Gyr) at larger distances. Older stellar populations dominate in the inner 250~pc.

\hfill{\bf Keywords}: galaxies: individual (Mrk\,573) -- galaxies: Seyfert -- infrared: galaxies -- galaxies: stellar population

\end{abstract}

\section{Introduction}

Stellar population (SP) synthesis using multi-wavelength spectra has being used to constrain the star formation history (SFH) of host galaxies of active galactic nuclei (AGN), looking in particular for the presence of recent star formation close to the AGN, supporting the so-called AGN-Starburst connection \citep{perry-dyson85,terlevich-melnick85,norman88}. Indeed, previous studies have shown that massive star-forming regions are commonly detected in the inner kiloparsec of active galaxies \citep{imanishi2000,thaisa2000,imanishi2002,ardila2003,rogerio07,dors08,rogerio09a}. Both, nuclear activity and star formation can be fed by gas inflows towards the nucleus, providing a gas reservoir in the central region of the galaxy. Indeed, inflows of gas have been observed using optical and near-infrared (near-IR) integral field spectroscopy  of nearby galaxies \citep[e.g.][]{fathi2006,riffel2008,fischer15}.  This feeding process is associated with the presence of nuclear bars or non-axis-symmetric features, spiral arms, or tidal interactions \citep{knapen2000,mw02,mw04a,mw04b}.

The SFH of galaxies can be used as a constraint for their formation and evolution, being a fundamental ingredient of theoretical models. Many studies of galaxy evolution in the infrared spectral range are strongly based on Evolutionary Population Synthesis (EPS) models \citep{capozzi16}. The main parameters of the stellar populations (SPs), as their ages, SFHs and stellar masses are derived through the EPS models. However, these models are still being refined and of particular importance is the contribution of Thermally Pulsing Asymptotic Giant Branch (TP-AGB) stars that play an important role in defining the shape of the spectra in the near-IR wavelengths \citep[e.g.][]{maraston05,marigo08,salaris14,rogerio15}. Construct EPS models for evolved stars, such as TP-AGB stars, is not an easy task due to  their complex inner structures, convection, and the eventual mass ejections.

Recently, SP studies of nearby Seyfert galaxies using near-IR spectroscopy have become more frequent \citep[e.g.][and references therein]{rogerio09a}. The near-IR spectral range, besides allowing us to access regions highly obscured by dust, also hosts spectral fingerprints originated in massive and evolved stars, as those found in the Red Supergiant (RSG) and TP-AGB phases. These phases are responsible for a large fraction of the near-IR stellar continuum and can be used as tracers of young to intermediate SPs with ages between 200\,Myr and 2\,Gyr \citep{maraston05,rogerio07,rogerio08,salaris14,rogerio15}. Additionally, in this spectral range, one can detect hot dust emission associated to the putative torus surrounding the central AGN, and some contribution also from the AGN featureless continuum (FC) emission, originated in the accretion disk \citep{riffel2009b}.  The detection and characterization of these components is fundamental to understand the AGN spectral energy distribution and investigate the impact of the AGN in the evolution of its host galaxy.

For the reasons pointed out above, near-IR spectroscopy is a powerful tool to investigate both the stellar populations and the unresolved emission from the dusty torus and accretion disk surrounding the supermassive black hole (SMBH). With the aid of the integral field capability at 8-10 meter telescopes, it is also possible to map the spatial distribution of the stellar population in the circumnuclear region of nearby active galaxies, a study that our group -- 
AGNIFS --  has been doing using the instrument NIFS (Near-Infrared Integral Field Spectrograph) at the Gemini North Telescope.

In a recent work, we have used near-IR integral field spectroscopy and optical long-slit spectra to map the emission-line and stellar kinematics of the inner $700 \times 2100$ pc$^2$ of Mrk\,573 using the NIFS at Gemini North and Dual Imaging Spectrograph (DIS) at Apache Point Observatory, respectively \citep{fischer17}. From this work, we found that flux distributions of ionized and molecular gas, while distinctly different, were morphologically related as arcs of molecular \H2 gas connected ionized [S\,{\sc iii}] gas features from outside the NLR bicone. We also found that molecular gas kinematics outside the NLR, and ionized gas kinematics at great distances from the nucleus in the extended NLR (ENLR), show signatures of rotation as observed from our stellar kinematics analysis. These observations suggest that the ionized gas kinematics and morphology in Mrk\,573 can largely be attributed to material originating in the rotating disk of the host galaxy. Deviations from pure rotation were observed along the NLR projected axis at radii r $<$ 750 pc and interpreted as being due to the radiative acceleration of material in the host disk. As the radiatively accelerated gas in the host disk goes to distances smaller than the length of the full NLR/ENLR, we concluded that these outflows may have a smaller range of impact than previously expected. The host disk galaxy presents an inclination of\, $i = 43^{\circ}$, with the north edge corresponding to the side nearest us \citep{fischer17}.

The stellar content of the central region of Mrk~573 was studied by  \citet{raimann03} using long-slit spectra obtained with the 4-m Mayall telescope of Kitt Peak National Observatory. They oriented the 1\farcs5 width slit along the PA=161$^{\circ}$ and mapped the stellar populations in the inner 8$^{\prime\prime}$ (2.7 kpc). They found that the flux at 4020\,\AA\ is dominated by an old SP component. Within the inner 2$^{\prime\prime}$ up to 70\,\% of the flux is due to emission of 10~Gyr SPs, while the contribution of these populations decreases at larger distances, being responsible for about 40\,\% of the observed flux at 8$^{\prime\prime}$ from the nucleus.  Intermediate-age (1~Gyr) SPs contribute with about 20\,\% of the nuclear flux, and their contribution increases to at larger distances, reaching 60\,\% outwards. The contribution of younger SPs is very small at all locations. The SP reddening values are in the range  $E(B-V)=0-0.4$, with the highest values seen at 2$^{\prime\prime}$ southeast of the nucleus.  On the other hand, \citet{rogerio09a} found that the near-IR continuum in the inner 0\farcs8$\times$1\farcs6 of Mrk\,573 is dominated by intermediate-age populations with ages ranging from 100\,Myr to 2~Gyr. The contribution of these populations reach 53\,\% of the flux at 1.2\,$\mu$m and is diluted by a featureless continuum (FC) component, which contributes with 22\,\% of the continuum. \citet{ramos2009} found also that intermediate age stars dominate the near-IR nuclear continuum, but they did not found an evidence for the FC component.

 This paper is organized as follows: in Section 2 we discuss the observations and data reduction procedures,  Section 3 shows maps of the flux and mass-weighted contribution of each SP,  which are discussed  in Section 4. Section 5 summarizes the conclusions of this work.

\section{Observations, data reduction and analysis}\label{obs}

In this paper we use the spectral synthesis technique to map the ages of the stellar populations of the inner 500\,pc radius of the Seyfert~2 galaxy Mrk\,573. 
Mrk\,573 is a nearly face-on, early-type galaxy, morphologically classified as RSAB(rs)$^+$ \citep{vaucoleurs91}, presenting a bright extended emission line region  ($\sim1$\,kpc) along the direction of its radio emission \citep{unger87,haniff88} and high-ionization emission-lines \citep{thaisa96}. Three radio continuum sources were first detected by \citet{ulvestad84}, one in the nucleus of the galaxy and the two radio lobes along position angle  $PA=122^{\circ}$.  It harbors a Seyfert 2 nucleus \citep{tsvetanov92} and is located at a distance of $\sim73$\,Mpc  \citep{springob05}, for which 1$^{\prime\prime}$ corresponds to 350 pc at the galaxy.

Z, J and K band integral-field spectroscopy (IFS) of Mrk\,573 have been obtained with the Gemini-north Near-Infrared Integral-Field Spectrograph \citep[NIFS--][]{mcgregor03} operating with Gemini North Adaptive Optics system ALTAIR. Observations of Mrk\,573 were obtained under the Gemini programme GN-2010B-Q-8 (PI: Michael Crenshaw) in 2010B and 2011A semesters, following the standard Object-Sky-Object dither sequence.
Six exposures of 600s each were performed in the K-band, centred at $2.3\,\mu$m and covering the spectral range from $2.1\,\mu$m to $2.5\,\mu$m, 5 exposures of 600\,s for J-band, centred at $1.25\,\mu$m, covering the spectral region from $1.1\,\mu$m to $1.3\,\mu$m and 6 exposures of 500\,s for Z-band, centred at $1.05\,\mu$m and  covering the spectral region from $0.94\,\mu$m to $1.14\,\mu$m. 

The NIFS has a square field of view of $3\farcs0\times3\farcs0$, divided into 29 slices with an angular sampling of 0$\farcs$10$\times$0$\farcs$04, and was oriented along the position angle PA$=133^{\circ}$, measured relative to the orientation of the slices. 

The data reduction was accomplished using tasks contained in the {\sc nifs} package, which are part of {\sc gemini iraf} package, as well as standard {\sc iraf} tasks and Interactive Data Language (IDL) routines. The process followed the standard procedure of near-IR spectroscopic data reduction,  including trimming of the images,  flat-fielding, sky subtraction, wavelength and s-distortion calibrations. The telluric absorption bands were removed by dividing the spectra of the galaxy by a normalized spectrum of a telluric standard star, observed just before and/or after the galaxy exposures. The final spectra were then flux calibrated by interpolating a black-body function to the spectrum of the telluric standard star. Individual exposure datacubes were created with an angular sampling of 0\farcs05$\times$0\farcs05, which were combined to obtain a single cube for each band, using the nucleus of the galaxy as reference for the astrometry. The final data cubes cover the inner  $\approx3\farcs0\times3\farcs0$ of Mrk\,573, corresponding to $\sim$1$\times$1\,kpc$^2$ at the galaxy.

The spatial resolution is $\sim45$\,pc for the J and K bands, as estimated from the full width at half maximum (FWHM) of the brightness profile of the telluric standard star, while for the Z band the performance of ALTAIR is worse and the resulting spatial resolution is about 55\,pc.  The spectral resolution is $\sim50$\kms\ for all bands, as obtained from the typical FWHM of arc lamp lines.

Since the performance of the adaptive optics at the Z band is worse, compared with the J and K bands,  and the signal-to-noise ratio of the Z band continuum spectra is also lower, we used only the J and K band datacubes to perform the spectral synthesis. These cubes were combined to a single datacube with a constant spectral bin of 5\,\AA\ and 0\farcs15$\times$0\farcs15 angular sampling. More details about the observations and data reduction procedure can be found in \citet{fischer17}.

\begin{figure*}
\centering
\includegraphics[scale=0.8]{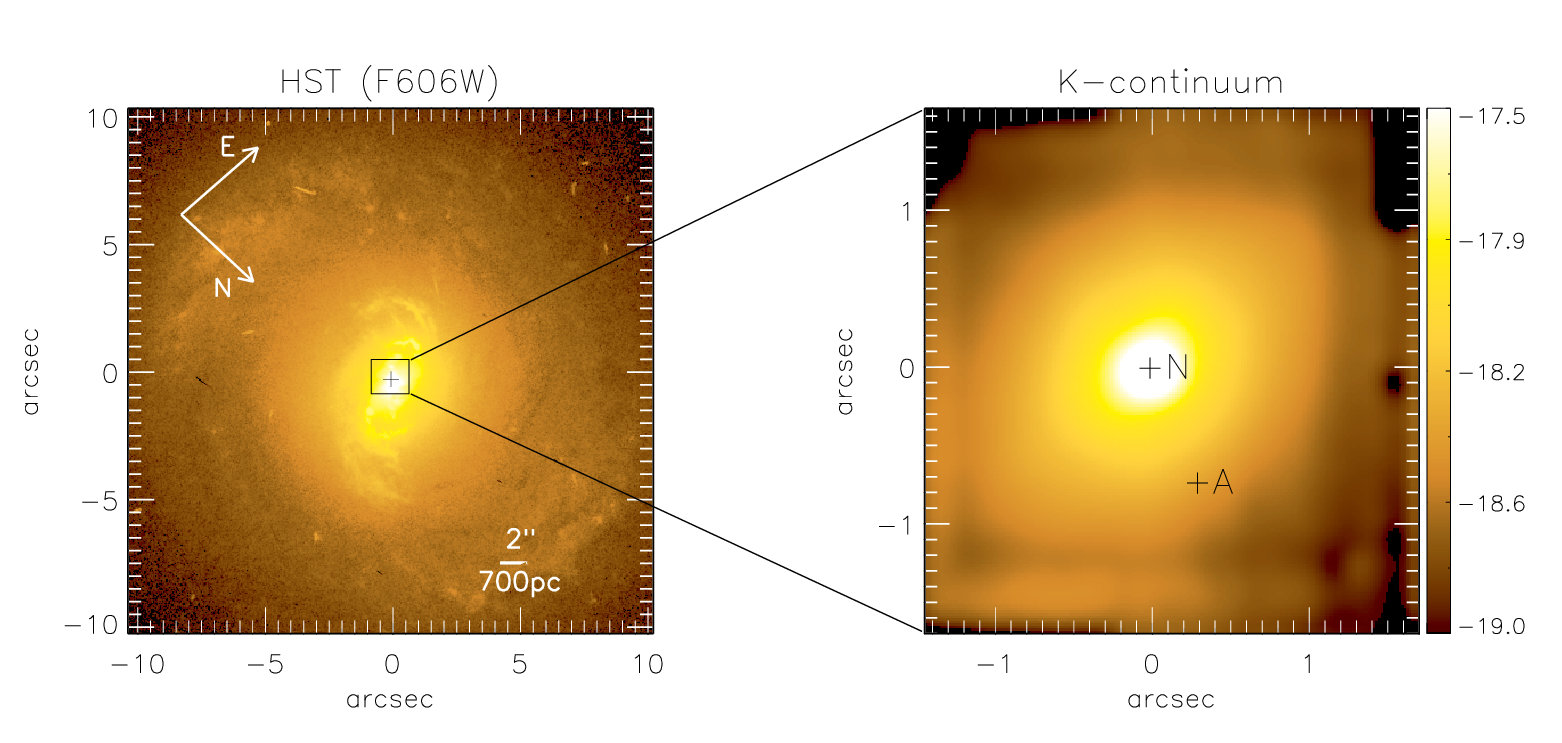} \quad
\includegraphics[scale=0.5]{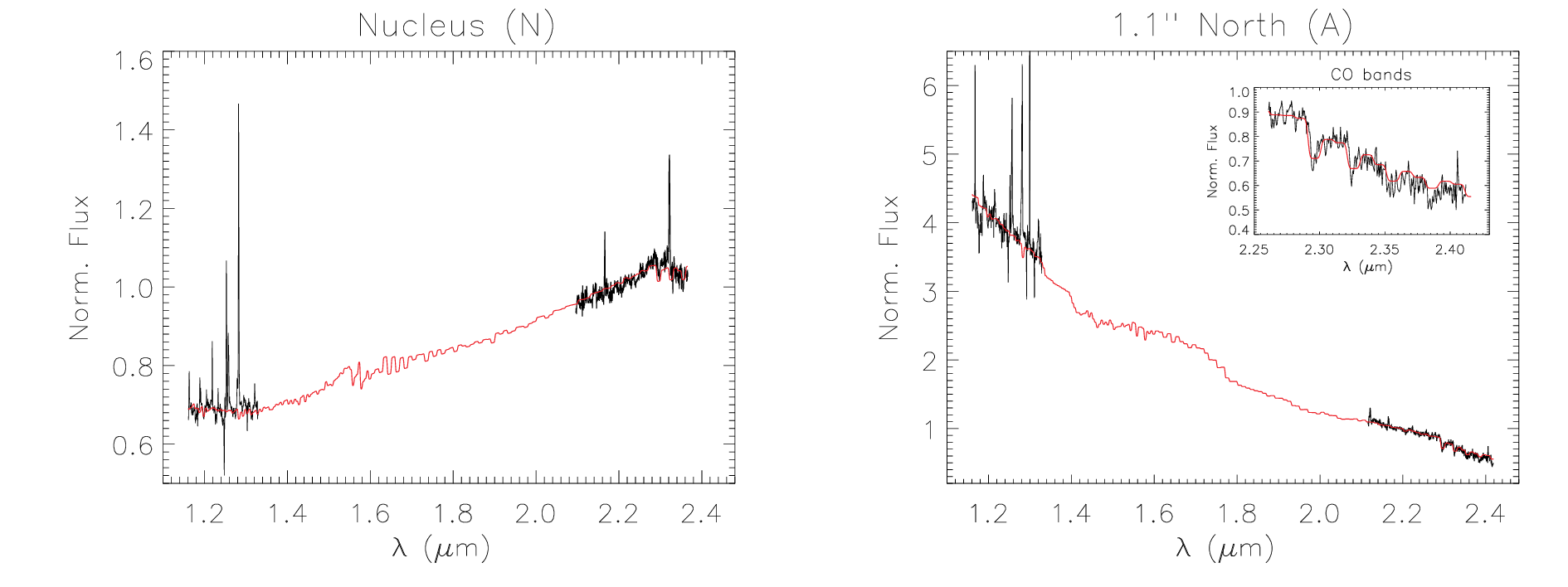} 
\caption{\small The left panel shows an optical HST image of Mrk\,573 through the filter F606W, with the NIFS FOV indicated by the central square. The right panel displays a continuum image in the K-band, in logarithmic flux units per pixel. The central cross on the panels
indicate the position of the nucleus. The bottom panels show the synthesis results (red) for two spectra (black): the nucleus and position A identified in the top right panel. The flux was normalized at 21955\,\AA\, and emission lines were masked. The HST image has been rotated to the same orientation of NIFS data (PA$=133^{\circ}$), indicated in the top-left corner of the top-left panel.
}
\label{hst573}
\end{figure*}

\subsection{Spectral synthesis}

The integrated spectrum of an active galaxy comprises a set of components such as stellar, gas, dust, as well as nuclear components such as a black-body (from the torus) and power-law emission from the accretion disk. One way to disentangle these components is by the technique of spectral synthesis, which allows to quantify the contributions of each one these components to the spectrum.

A widely used code is the {\sc starlight} \citep{asari07,cid04,cid05a,cid05b}, which fits the continuum searching for the best description of the observed spectrum by reproducing it with different proportions of the supposed components that sum up to the observed spectrum. These components comprise the base of spectral elements. In this way,  the ``key" of the spectral synthesis technique is to provide a base of elements including all possible components observed in galaxies \citep{cid04,cid05a,cid05b,riffel2009a}. When fitting near-IR spectral data one needs to have in mind that this region host characteristic absorption features, being the most common the  CN, CO, VO, ZrO and TiO absorptions bands, which are attributed to evolved stars, as those in the RGB and TP-AGB phases \citep[e.g.][]{maraston05,rogerio07,rogerio15}. Thus, the simple stellar population (SSP) models used to fit near-IR data need to include these features. Therefore, we selected the \citet{maraston05} SSP models, that include empirical data for the the TP-AGB evolutionary phase. 

The set of spectral elements used here is composed by the SSP models of \citet{maraston05} and are described in \citet{rogerio09a}. In short, they include 12 ages ($t=$ 0.01, 0.03, 0.05, 0.1, 0.2, 0.5, 0.7, 1, 2, 5, 9, 13 Gyr) and 4 metallicities
(Z $=0.02, 0.5, 1, 2\,$Z$_\odot$). We also included black-body functions for temperatures in the range $700-1400$ K in steps of 100 K and a power-law ($F_{\lambda} \sim \lambda^{-0.5}$), in order to account for possible contributions from hot dust emission and from a featureless continuum (FC) in the nucleus of the galaxy. For details see \citet{riffel2009a}.

The fit is carried out in {\sc starlight} by minimizing the following equation:

\begin{equation}
\chi^2=\sum_{\lambda}[(O_\lambda - M_\lambda)\omega_\lambda]^2,
\label{minimum}
\end{equation}
where $O_\lambda$ is the observed spectrum, $M_\lambda$ is the fitted model,
$\omega_\lambda = 1/e_\lambda$  and $e_\lambda$ corresponds to the associated uncertainties to the observed spectrum. 

The emission lines and spurious data were excluded from the fit by setting their weight as zero. Each model spectrum is obtained by:

\begin{equation}
 M_\lambda=M_{\lambda0}\left[\sum_{j=1}^{N_\star}x_j b_{j,\lambda} r_\lambda \right] \otimes G(\upsilon_\star,\sigma_\star),
\label{model-spectrum}
\end{equation}
where $M_{\lambda0}$ is the synthetic flux at the normalization wavelength free of any emission or absorption line, $\vec{x}$ is the population vector, whose components $x_j$ ($j=1,$...$,N_\star$) represent the fractional contribution of each SSP in the spectral base, $b_{j,\lambda}$ is the normalized spectrum of the $j$th SSP component of the base, $\otimes$ corresponds the convolution operator and $G(\upsilon_\star,\sigma_*)$ is the Gaussian distribution used to model the line of sight velocity distribution, centered at velocity $\upsilon_\star$ with dispersion $\sigma_*$.  The extinction due to dust is modeled as an uniform screen following the extinction law of \citet{cardelli89}.

\begin{figure*}
\begin{center}
    \includegraphics[scale=0.75]{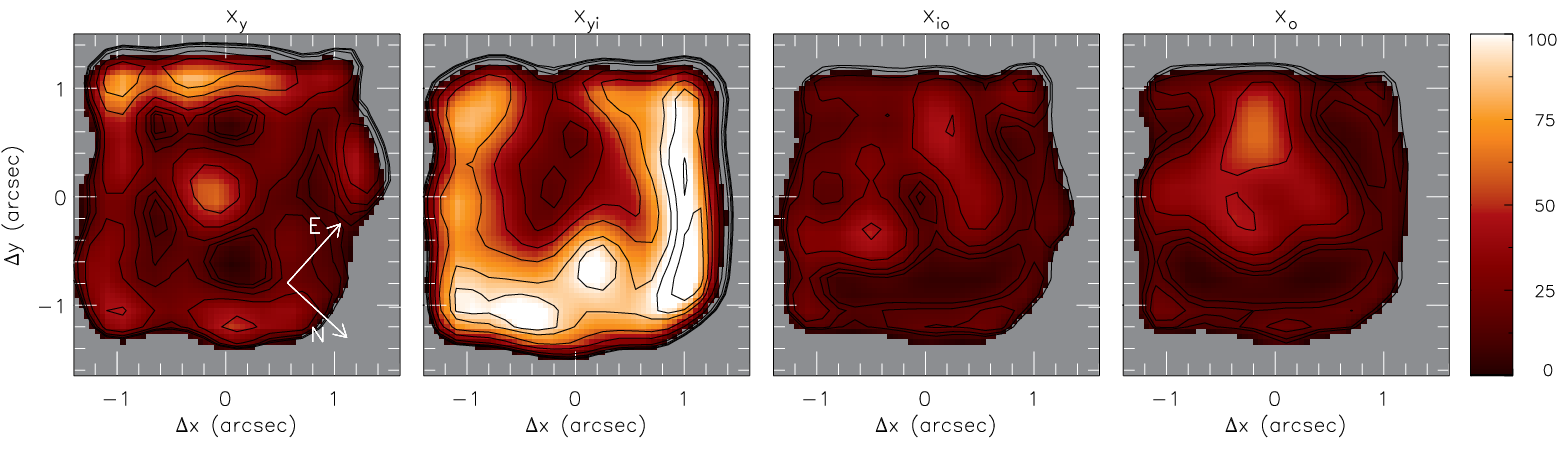} \qquad
    \includegraphics[scale=0.75]{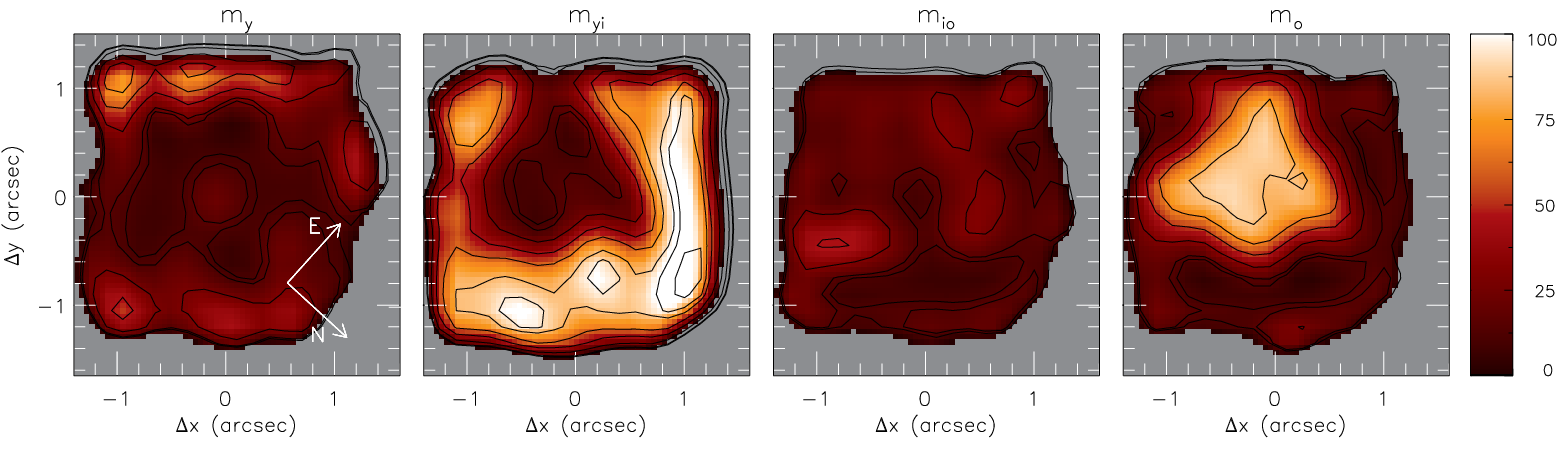}
\caption{\small From left to right we show the distributions of: in the top row, percent contributions to the 2.2$\mu$m continuum of young (\xy $\leq 100$\,Myr), young-intermediate ($ 100 < $ \xyi $\leq 700$\,Myr), intermediate-old ($700\, {\rm Myr} < $ \xio $ \leq 2$\,Gyr) and old ($2 < $ \xo $\leq 13$\,Gyr) age components; In the bottom row, we show the corresponding percent mass contributions (\my, \myi, \mio~and \mo). The spatial orientation (PA$=133^{\circ}$) is indicated in the left panels. Contours in the BB and FC panels correspond to 10 and 20\,\% flux contribution.}
\label{bri573}
\end{center}
\end{figure*}

\begin{figure*}
\begin{center}
    \includegraphics[scale=0.65]{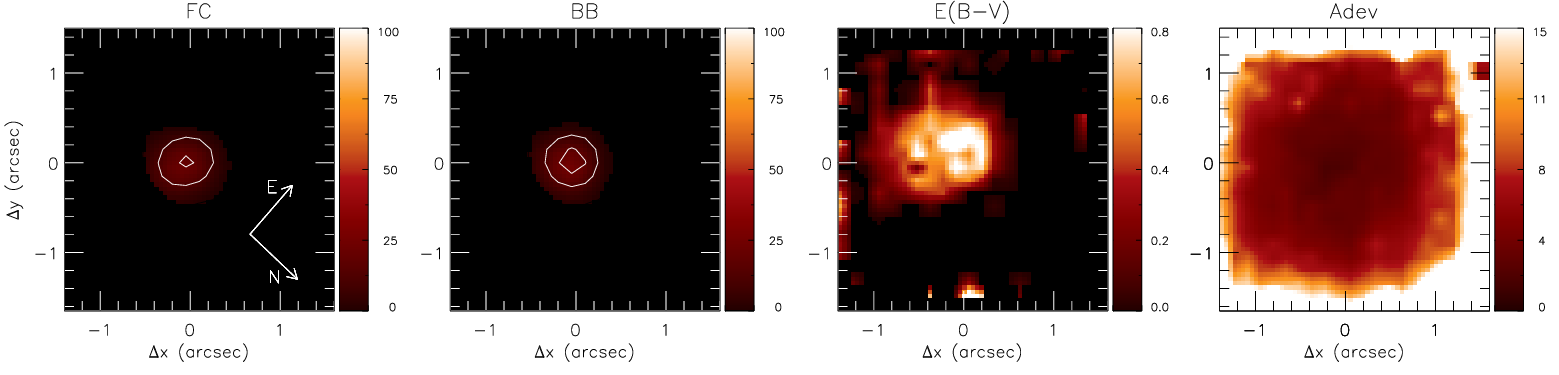}
\caption{\small From left to right: percent contributions to the 2.2$\mu$m continuum of the FC and BB emission, Adev and Av maps. The spatial orientation (PA$=133^{\circ}$) is indicated in the left panel.
}
\label{fc573}
\end{center}
\end{figure*}

\section{Results}\label{results}

An optical image of Mrk\,573 obtained with the Hubble Space Telescope (HST) Wide Field Planetary Camera 2 (WFPC2) through the filter F606W is shown in the top-left panel of Fig.\ref{hst573}. The central black square indicates the FOV of the NIFS observations. The right panel shows a continuum image of the nuclear region, in logarithmic flux units per pixel, acquired from the NIFS datacube in the spectral range between 2.25 and 2.28 $\mu$m, without emission or absorption lines. Spectra from two positions (N and A), extracted within an aperture of 0\farcs15$\times$0\farcs15, are shown in the bottom panels. The central  ``cross'' corresponds to the location of the nucleus, which was defined as the peak of continuum emission. 

The bottom panels of Fig.\ref{hst573} show the results of the spectral synthesis overploted to the observed spectrum (black), for the nucleus (N) and for position A, indicated on the continuum map (top-right panel). In the spectral synthesis, we masked out emission lines and spurious features, fitting only the regions with stellar absorption features and featureless continua.

The fits for all spaxels are very similar to those shown in Fig.~\ref{hst573}. The quality of the fits can be evaluated using the rightmost panel of Fig.\ref{fc573}, where we show the map of the mean percent deviation over all fitted pixels (Adev=$|O_{\lambda} - M_{\lambda}|/O_{\lambda})$.

Since small stellar population differences in the spectra are washed away due to the uncertainties on the observations, we have followed \citet{cid04} and binned the contribution of the individual SSP, $x_j$, into a coarser population vector as follows \citep{riffel2010,rogerio09a}:
{\bf young} ($x_y$: $t \le 100\,$Myr); 
{\bf young-intermediate} ($x_{yi}$: $100 < t \le 700\,$Myr);
{\bf intermediate-old} (\xio: $700$\,Myr $ < t \le 2\,$Gyr) and
{\bf old} (\xo: $2 < t \le 15\,$Gyr).

Fig.~\ref{bri573} (top) shows the results of the spectral synthesis in maps of the main stellar population components (SPCs) percent contributions to the 2.2$\mu$m continuum light within inner $\sim$500\,pc of Mrk\,573. Grey regions in these maps correspond to masked locations where we were not able to get good fits, with $adev > 15$. The \xy\ map shows that the young population contributes with up to 50\% of the continuum within $\sim$0\farcs5, with the highest contribution seen at 0\farcs3 south of the nucleus. It decreases outwards and then increases again at $\sim1^{\prime\prime}$ from the nucleus in a partial ring structure showing values of up to 100\% to the south and south-east. The rest of the ring is dominated by the contribution from the \xyi\ population that reaches up to 100\% at the ring.
The last two panels (\xio\, and \xo) show a flux contribution of old SPs in the central region of up to  60\% mostly inside the ring.

Once the SSPs are in the form $L_\odot\AA^{-1} M_\odot^{-1}$, thus a light-to-mass-ratio spectrum, the {\starlight} code computes the mass fractions based on the $L/M$ ratio\footnote{More details can be found in the STARLIGHT User Guide at www.starlight.ufsc.br.}. The maps with the mass fractions for each binned age group are shown in the bottom panels of Fig.~\ref{bri573}. The spatial distribution of each mass contribution, as expected, is similar to that observed to the light fractions, however, due to the nonlinear $M/L$ relation, the \mo\ maps show higher contributions in mass than in flux, with values of up to 95\%.

In Fig.~\ref{fc573} we present the maps for the FC and BB components that contribute to the observed continuum emission within 0\farcs45 (180\,pc). Although by the map it seems that both have the same value, in fact the FC has a contribution of up to 35\% and the BB a contribution of up to 50\%, in the central pixels. In the same figure we show the $E(B-V)$ reddening map, which displays values of up to $0.8$ and the $Adev$ map with most of the values $\lesssim$10\%.

\section{Discussion}\label{discussion}

\subsection{Distribution of the stellar populations}
Studies of the stellar populations in nearby Seyfert galaxies based on near-IR long-slit spectra show a substantial fraction of  intermediate-age SPs in the inner few hundreds of parsecs \citep[e.g.][]{rogerio09a}, while at optical wavelengths signatures of recent episodes of star formation are seen for about 40-50\,\% of nearby bright Seyfert~2 galaxies at similar spatial scales \citep[e.g.][]{sb01,raimann03,cid04,sarzi07}. Young and intermediate age stellar populations in the inner few hundreds of parsecs of Seyfert galaxies are also detected in recent near-IR IFS studies \citep[e.g.][]{davies07,riffel2010,rogerio11b,ngc1068}. In addition, a correlation between the distribution of intermediate-age stars with low-stellar velocity dispersion ($\sigma_*\sim50-70$\,\kms) rings has been observed for some objects \citep{riffel2010,rogerio11b}, indicating that these low-$\sigma_*$ rings are originated from SPs that still preserve the ``cold" kinematics of the gas they were formed (being thus younger than the surroundings), as claimed to explain the $\sigma_*$-drops commonly reported for more than 20 years \citep[e.g.][]{emsellem01,marquez03}.

Our results for Mrk\,573 show that old SPs are dominant in the inner 0\farcs8 (300\,pc), while at larger distances from the nucleus (up to $\approx$1\farcs2 -- covered by the FOV of our observations)  young-intermediate SPs dominates the stellar mass and K-band continuum emission (see Fig.~\ref{bri573}).  At optical wavelengths, \citet{schmitt99} found that 82\% of the nuclear continuum at 5870\,\AA\ of Mrk\,573 is due to a 10\,Gyr stellar population. They used an integrated spectrum within an aperture of $2^{\prime\prime}\times 2^{\prime\prime}$ and their result is in agreement with other studies of the stellar populations in optical wavelengths \citep{raimann03,sb01}. On the other hand, near-IR spectral synthesis of the nucleus for an aperture of 0\farcs8$\times$1\farcs6 points out that intermediate-age stars contribute to 53~\% of the K-band flux \citep{riffel2009a}, being also in agreement with results obtained by \citet{ramos2009}, who found that the nuclear H and K band emission is dominated by late-type giants with ages between 100~Myr and 1~Gyr.

\begin{figure}
\begin{center}
    \includegraphics[scale=0.51]{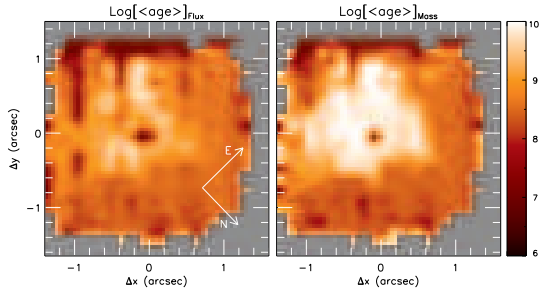}
\caption{Logarithm of the mean age weighted by flux (left) and stellar mass (right).
}
\label{agemetal573}
\end{center}
\end{figure}

The simplest way to represent the mixture of stellar populations of a galaxy is estimating its mean light ($<{\rm log}t_{\star}>_L$) and mass ($<{\rm log}t_{\star}>_M$) weighted stellar age. Following \citet{cid05b}, 

\begin{equation}
 <{\rm log}t_{\star}>_L = \sum^{N}_{J=1}x_j{\rm log}t_j
\end{equation}
and
\begin{equation}
 <{\rm log}t_{\star}>_M = \sum^{N}_{J=1}m_j{\rm log}t_j.
\end{equation}
While the former is more representative of younger ages, the latter is enhanced by the old SPC \citep{rogerio09a}. In Fig.~\ref{agemetal573} we show the maps for the mean age light- (left panel) and mass- (right panel) weighted in logarithmic units (years).
The mean light- and mass-weighted ages over whole field of view are $<{\rm log}t_{\star}>_L=8.39$ and $<{\rm log}t_{\star}>_M = 8.61$, respectively. These values are also in good agreement  with those found in \citet{rogerio09a} for the nuclear 0\farcs8$\times$1\farcs6.

 As discussed above, studies based on seeing limited observations found that old SPs dominate the emission at optical bands, while in the near-IR intermediate-age stars are dominant \citep{schmitt99,sb01,rogerio09a,ramos2009}. These studies are based on measurements of a nuclear spectrum that actually integrates the light within a few hundreds pc of the nucleus, what is comparable to the whole NIFS FOV. The NIFS adaptive optics observations allowed us to spatially resolve the distribution of the stellar populations in the inner 600~pc of Mrk~573 at a spatial resolution of $\sim50$~pc, at least 5 times better than that of the previous long-slit observations.  We have shown that recent ($t<700$\,Myr) star formation dominates at distances larger than 300~pc from the nucleus (at least up to $\approx$500\,pc from the nucleus), while at smaller distances older stars dominate the near-IR continuum emission and the stellar mass content of Mrk\.573.

\subsection{Spatial correlation among stellar populations, velocity dispersion and H$_2$ emission}

\begin{figure}
\begin{center}			
    \includegraphics[scale=0.42]{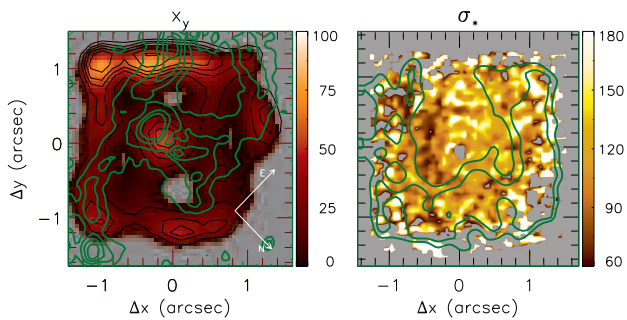} 
\caption{Map of the young SPC with green contours overlaid showing the \H2$\lambda2.12\,\mu$m emission-line flux distribution (left panel), and stellar velocity dispersion map (right panel) with overlaid contours of the young-intermediate SPCs distributions in green. 
}
\label{pops}
\end{center}
\end{figure}

Our group AGNIFS (AGN Integral Field Spectroscopy)  has started to characterize the stellar population of the inner kiloparsec of galaxies using the  {\sc starlight} code \citep[e.g.][]{cid04,cid05a,cid05b} via spectral synthesis of IFS obtained with the Gemini Near-infrared Integral Field Spectrograph \citep[NIFS-][]{mcgregor03}. To date, we studied four nearby Seyfert galaxies (Mrk\,1066, Mrk\,1157, NGC\,1068 and NGC\,5548). For NGC\,1068, we found two episodes of recent star formation: one at 300\,Myr ago, extending over the inner 300\,pc of the galaxy and another at 30\,Myr ago, observed in a ring at $\sim$100\,pc from the nucleus and being associated to an expanding ring observed in warm H$_2$ gas emission \citep{ngc1068}. For Mrk\,1066 and Mrk\,1157, rings of intermediate age stars have been found, being correlated with low stellar velocity dispersion values ($\sigma_*\sim$50\kms), and interpreted as being originated by stars that still preserve the kinematics of the gas from which they formed \citep{riffel2010,rogerio11b}. 
In the case of NGC\,5548, the stellar population is dominated by an old ($>$2 Gyr) component between 160 and 300 pc from the nucleus, while closer to the nucleus, intermediate age stars (50~\,Myr--2\,Gyr) are dominant \citep{astor16}. Hot dust emission was detected for three galaxies (Mrk\,1066, NGC\,1068 and NGC\,5548), accounting for 30-90\,\% of the observed K-band nuclear flux, while for two galaxies the FC component was detected, contributing with $\sim$25\,\% of the K-band nuclear flux for NGC\,1068 and  $\sim$60\,\% for NGC\,5548 \citep{riffel2010,rogerio11b,ngc1068,astor16}.

In order to look for similar correlations for Mrk\,573, we present in the left panel of Fig.~\ref{pops} contours (in green) of the H$_2$\,$\lambda2.1218\,\mu$m flux distribution (left panel) overlaid on the young SPC distribution, while in the right panel we show the stellar velocity dispersion map with overlaid contours in green of the young-intermediate SPC. The H$_2$\,$\lambda2.1218\,\mu$m fluxes were measured by direct integration of the H$_2$ line profile from the datacube and subtracting the adjacent continuum. The $\sigma_*$  map was obtained by using the penalised pixel-fitting (pPXF) method of \citet{cappellari04} to fit the CO absorption band-heads at $\sim2.3\,\mu$m, using as templates spectra of late type stars from \citet{winge09}. The H$_2$ flux distribution shows two arc-shaped structures extended along the east-west, with the highest emission observed within a blob centred at 0\farcs3 south of the nucleus.

The  $\sigma_*$ map shows that most values range between 80 and 180\,\kms, with the lowest values being observed mainly to south, west and north-west of the nucleus at distances of 0\farcs5 from it.  More details about the molecular gas distribution and kinematics and stellar kinematics are presented in \citet{fischer17}. 

Fig.~\ref{pops} shows no clear correlation between the young SP distribution and the H$_2$ emission line map for Mrk\,573, except maybe at 0\farcs3 south of the nucleus where a blob of higher H$_2$ flux shows some overlap with the distribution of the young SPC around the nucleus. A similar trend is observed between low-$\sigma_*$ values and the distribution of the young-intermediate age SP, as shown by the green contours overploted to the sigma map. 
This support our previous similar findings for other active galaxies that the low-$\sigma_*$ structures are due to stars formed $\sim$ 100--700\,Myr ago, that still are not in orbital equilibrium with the bulge stars and possible were formed from gas recently accreted to the inner kiloparsec \citep[e.g.][]{barbosa06}.

\subsection{Radial distribution of the stellar populations in Mrk~573 and comparison with other galaxies}

\begin{figure*}
\centering
\includegraphics[scale=0.9]{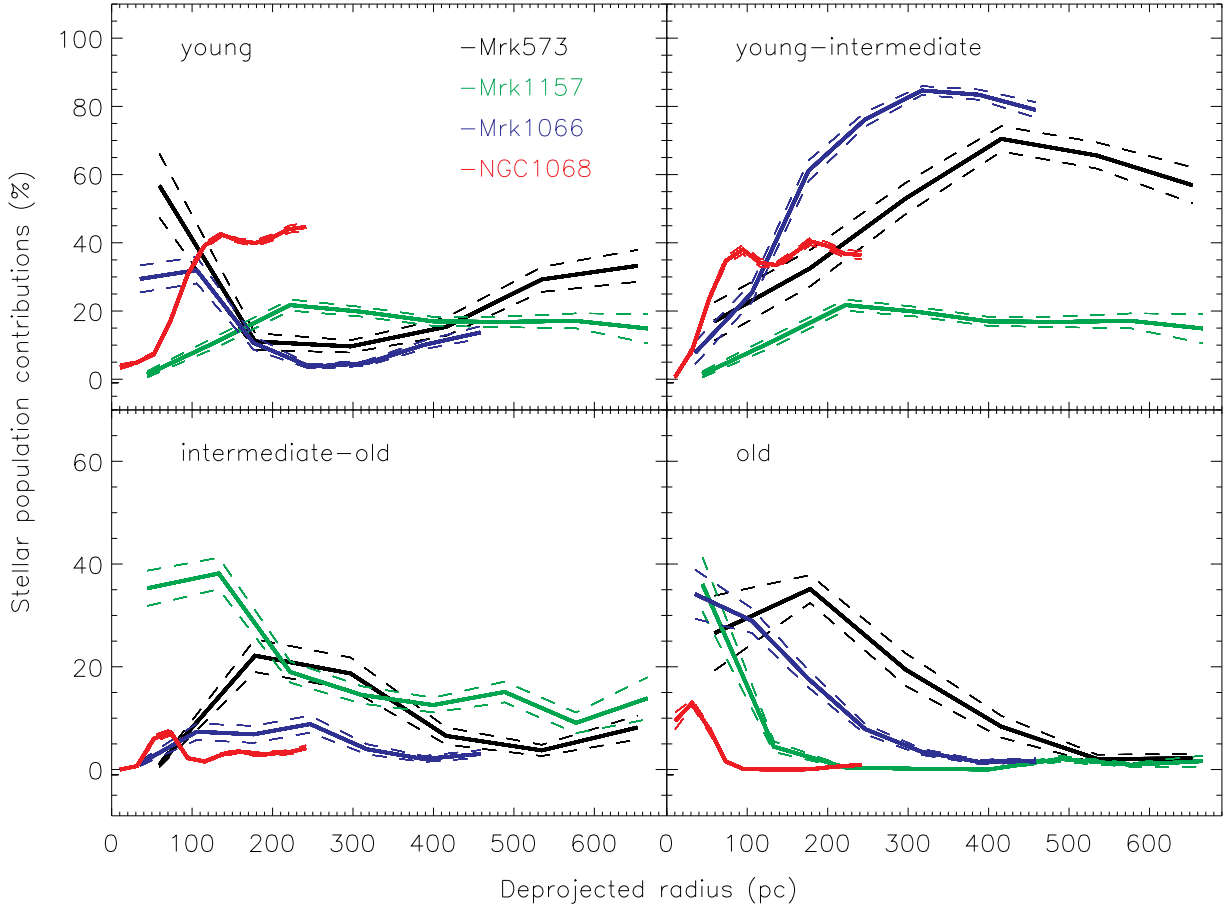}
\caption{Contribution to the flux at 2.2$\mu$m of each stellar population component as a function of the deprojected distance to the nucleus. Values for Mrk~573 are shown in black lines, for Mrk~1157 in green, for Mrk~1066 in blue and for NGC~1068 in red. The dashed lines show the standard deviation of the azimuthal mean at each radius. See text for more details.}
\label{radial_all_k}
\end{figure*}

In order to compare the radial distribution of stellar population contributions in Mrk\,573 with those of previous studies by our AGNIFS team for other active galaxies, using similar data and technique, we have built radial profiles of the young, young-intermediate, intermediate-old and old population contributions to the flux at 2.2~$\mu$m for all studied galaxies so far. The results are shown in Fig.~\ref{radial_all_k}. The age bins for all galaxies are the same (as described in Section~\ref{results}) and the derived values for Mrk~1066, Mrk\,1157 and NGC\,1068, are described in \citet{riffel2010}, \citet{rogerio11b} and \citet{ngc1068}, respectively. These plots were constructed by calculating the average contribution of each SPC to the flux at 2.2~$\mu$m within circular rings  with 0\farcs3 width in the (deprojected) plane of the disk of the galaxy. The orientation of the line of nodes ($\Psi_0$) and disk inclination ($i$) used in the deprojection were  obtained from the modeling of the stellar velocity field and presented in \citet{fischer17} for Mrk~573 ($\Psi_0=97^\circ$, $i=26^\circ$),  \citet{riffel2017} for Mrk~1066 ($\Psi_0=120^\circ$, $i=50^\circ$) and Mrk~1157  ($\Psi_0=114^\circ$, $i=45^\circ$)  and \citet{davies07} for NGC~1068 ($\Psi_0=85^\circ$, $i=40^\circ$).
 
Although contribution of young stellar populations of at least 20\% are observed at the nucleus only for Mrk\,573 and Mrk\,1066, they are present for all studied galaxies within the inner 200--300 pc, in good agreement with results obtained from optical \citep[e.g.][]{sb01,sarzi07} and near-IR studies \citep[e.g.][]{davies07,riffel2009a}. These young stars possibly originate from a reservoir of gas recently accumulated in the central region of active galaxies. One possibility is that these reservoirs have been built by gas streaming motions along spiral arms and nuclear bars, seem at similar scales in many active galaxies \citep[e.g.][]{fathi2006,ms09,diniz15}. The nuclear activity may also have been triggered by the presence of this gas reservoir, due to gas directly accreted by the SMBH or to accretion of gas ejected by the recently formed young stars \citep[e.g.][]{davies07,riffel2009a}.

Some contribution of young SPs is also observed at distances larger than 500\,pc from the nucleus, while the young-intermediate age SPs show their highest contribution at distances of 300--500\,pc from it. The intermediate-old population is observed mainly within the inner 400~pc (with its highest contribution at $\sim$200--300 pc from the nucleus) and old populations are mainly observed within the inner 250~pc. The results for the oldest SPCs are in agreement with the inside out star formation scenario, in which a gradient of stellar populations ages is observed, with old stellar populations seen at the nucleus and with decreasing ages outwards \citep{perez13,delgado2015,delgado2016}.

\subsection{Dust mass and AGN Featureless continuum}

We found that both the AGN FC (power-law) component and black-body emission contribute to up to 20\,\% of the observed K-band nuclear flux of Mrk~573, as seen in Fig.~\ref{fc573}. The nucleus of Mrk~573 is classified as Seyfert~2 and the inclusion of these components is necessary to fit the nuclear spectrum of at least 25\,\% of Seyfert 2 galaxies, while more than 50\,\% of Seyfert 1 galaxies show FC and hot dust contribution \citep{rogerio09a}. The detection of the power-law component for Mrk~573 suggests that radiation from the accretion disk is coming out through the torus, in good agreement with the detection of an obscured Narrow-Line Seyfert 1 nucleus as indicated by the presence of a broad component in the Pa$\beta$ emission line \citep{ramos2008}.

The spectral synthesis confirmed the presence of an unresolved black-body component at the nucleus, previously detected in the infrared \citep{ramos2009,schlesinger09,rogerio09a}, and possibly due to the dusty torus surrounding the SMBH. We have estimated the mass of the hot dust following \citet{riffel2009a} and using the formalism of \citet{barva1987}, for dust composed by grains of graphite.

The IR spectral luminosity of each dust grain, in erg\,s$^{-1}$\,Hz$^{-1}$, 
can be written as 
\begin{equation}
 L^{\rm gr}_{\nu,{\rm ir}} = 4\pi^2 a^2 Q_\nu B_\nu(T_{\rm gr}),
\label{lumigrain}
\end{equation}
where $a = 0.05\,\mu$m is the grain radius, $Q_\nu = 1.4 \times 10^{-24}\nu^{1.6}$
is its absorption efficiency and
$B_\nu(T_{\rm gr})$ is its spectral distribution assumed to be a Planck function
for a temperature $T_{\rm gr}$.

The total number of graphite grains can be obtained from
\begin{equation}
 N_{\rm HD} \sim \frac{L^{\rm HD}_{\rm ir}}{L^{\rm gr}_{\rm ir}},
\end{equation}
where $L^{\rm HD}_{\rm ir}$ is the total luminosity of the hot dust,
obtained by integrating the flux of each black-body component contribution from the synthesis. Then, we multiplied the integrated normalized flux by the normalization flux at 21955\AA\, and convert it to the adequate units
(from erg\,s$^{-1}$\,cm$^{-2}$\,\AA$^{-1}$ to erg\,s$^{-1}$\,Hz$^{-1}$).  In order to obtain $L^{\rm gr}_{\rm ir}$, we have integrated the Eq.~\ref{lumigrain}
for all temperatures, ranging them from 700 to 1400\,K, in steps of 100\,K.

Finally, the hot dust mass can be obtained by the equation \citep[e.g.][]{ardila2005}:

\begin{equation}
 M_{\rm HD} \sim \frac{4\pi}{3}a^3N_{\rm HD}\rho_{\rm gr},
\end{equation}
where $\rho_{\rm gr} = 2.26$\,g\,cm$^{-3}$ is the density of the grain.
The total dust mass estimated for the nucleus of Mrk~573 by integrating over the entire field of view, is $M_{\rm HD}=1.3\times10^{-2}$\,M$_{\odot}$, which is within the range of masses observed for other active galaxies \citep[e.g.][]{rogerio09a,riffel2010,rogerio11b,ardila2006,ardila2005}.

\subsection{Extinction}

\begin{figure*}
\centering
\includegraphics[scale=0.35]{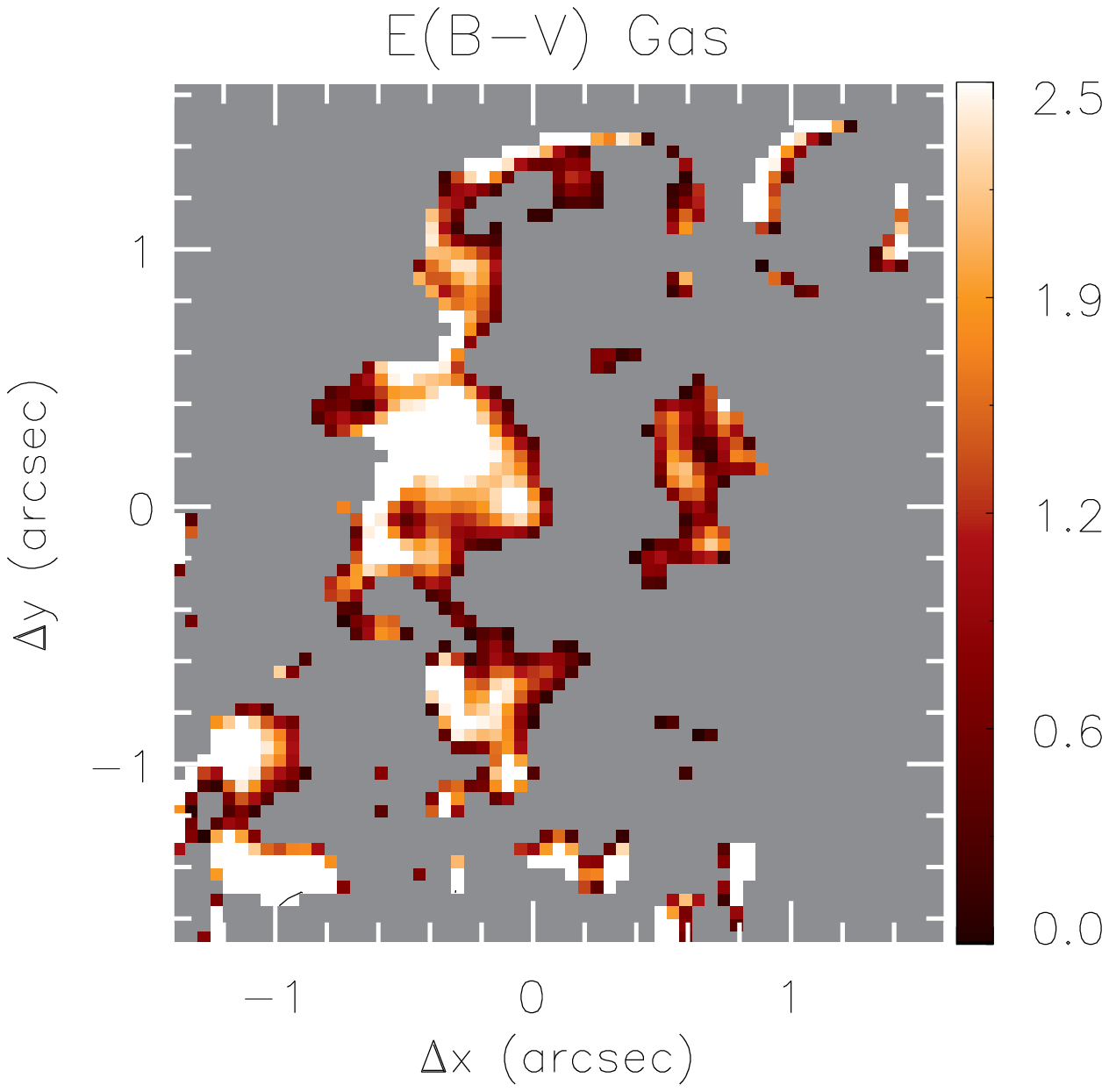} \qquad
\includegraphics[scale=0.33]{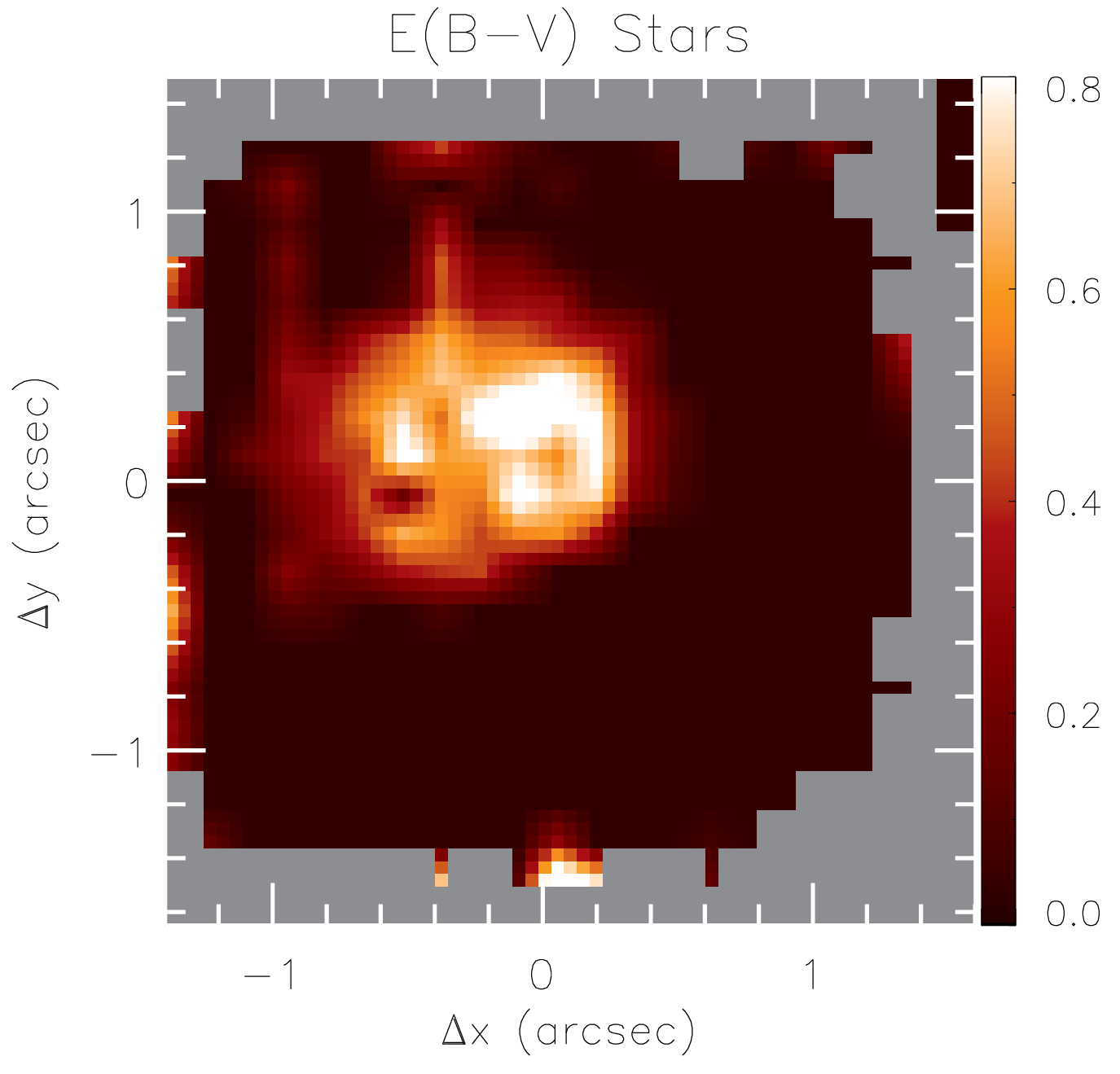} \qquad
\includegraphics[scale=0.33]{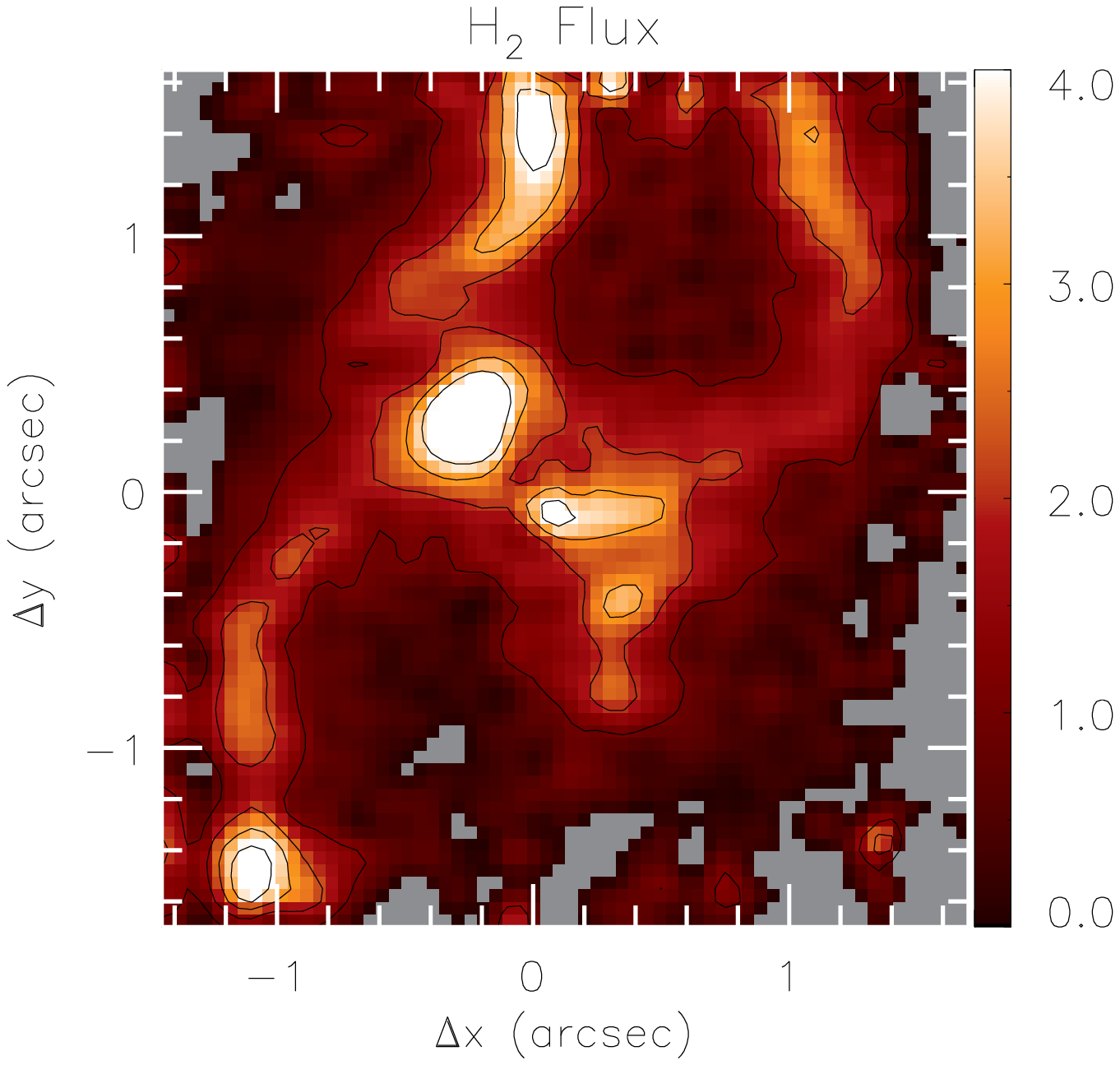} 
\caption{\small From left to right: $E(B-V)$ map for the gas from the line ratios between Pa$\beta$ and Br$\gamma$; stellar E(B-V) map obtained from \starlight~ and the H$_2\lambda$21218\AA~flux map.
}
\label{ebv}
\end{figure*}

The $E(B-V)$ map (Fig.~\ref{fc573}) obtained for the stellar population show higher values to the southwest side of the nucleus. We can compare this map with that for the gas extinction. A gas reddening map can be obtained by using the  Pa$\beta$/Br$\gamma$ emission line ratio via the following equation: 
\begin{equation}
 E(B-V)=4.74\,{\rm log}\left(\frac{5.88}{F_{Pa\beta}/F_{Br\gamma}}\right),
\end{equation}
where $F_{Pa\beta}$ and $F_{Br\gamma}$ are the fluxes of Pa$\beta$ and Br$\gamma$ emission lines, respectively. This equation was obtained using the  reddening 
law of \citet{cardelli89} and adopting the intrinsic ratio $F_{Pa\beta}/F_{Br\gamma}=5.88$ corresponding to case B recombination \citep{osterbrock06}. The Pa$\beta$ and Br$\gamma$ emission-line flux distributions were obtained by fitting the line profiles at each spaxel by Gaussian curves, and as discussed in \citet{fischer17} these lines show similar flux distributions to those of the [S\,{\sc iii}]$\lambda0.95\,\mu$m line.  The resulting  $E(B-V)$ map for the gas is shown in Fig.~\ref{ebv}.

The median value of the reddening for the gas over the whole field of view is $E(B-V)\sim2.5$ mag, considering only spaxels where both Br$\gamma$ and Pa$\beta$ emission lines were detected. For the stellar population we obtain a smaller value of $E(B-V)\sim$0.25 mag by averaging the map presented in Fig.~\ref{fc573}, showing that the gas extinction in the NLR is larger than that of the stellar population.   \citet{rogerio06} presented the nuclear spectrum of Mrk~573 for an aperture of 0\farcs8$\times$1\farcs6 obtained with the spectrograph SpeX covering the spectral range from 0.8 to 2.4\,$\mu$m. Using the their flux values for Br$\gamma$ and Pa$\beta$ emission lines, we obtain an  $E(B-V)$ value for the gas very similar to ours, while \citet{rogerio09a} obtained $E(B-V)\approx0.3$~mag for the stellar population using the same dataset.  A higher extinction for the gas is also observed at optical wavelengths, in which the optical depth of the continuum underlying the Balmer lines is about half of that for the emission lines \citep{calzetti94}. This difference may be due to the fact that most of the gas is located closer to the galaxy disk, where large amount of dust is expected,  while the near-IR continuum has an important contribution from stars of the bulge of the galaxy.

Comparing the gas and stellar $E(B-V)$ maps, we note that they show a similar distribution near the nucleus, with higher values observed to the southwest of the nucleus and values close to zero to the northeast of it. The observed extinction for the gas is larger than that of the stellar population, in good agreement with previous studies \citep[e.g.][]{rogerio08}. A higher extinction to the southwest of the nucleus is also supported by the dust lanes seen in the structure map presented in  Fig.~7 of \citet{fischer17}.

The rightmost panel of Fig.\,\ref{ebv} shows that the distribution of the hot molecular gas presents a good correlation with the gas $E(B-V)$, confirming the known association between the molecular gas and dust. The highest apparent concentration of hot molecular gas coincides with the region of highest extinction in the gas and in the stars at 0\farcs3 south of the nucleus, being also co-spatial with young stellar populations  (Fig.~\ref{bri573}).

\section{Conclusions}	

We used near-IR integral field spectra  at a spatial resolution of $\sim$50\,pc to map the stellar population distributions in the inner 500~pc of the Seyfert galaxy Mrk~573, as well as featureless continua contributions at the nucleus by combining the spectral synthesis technique with \citet{maraston05} SSP models. The main conclusions of this work are:

\begin{itemize}

\item Although the old stellar population ($x_i > 2$~Gyr) dominates the K-band continuum in the inner $\sim$250~pc (0\farcs75), within the $\sim$70 pc from the nucleus there is up to 30\% contribution from a young stellar population ($x_i < 100$~Myr). Beyond the inner $\sim$250~pc and up to the border of the FOV (500~pc), the young-intermediate age stellar populations (100--700~Myr) are dominant, in a structure resembling a partial ring where its contribution to the continuum reaches up to $\approx$\,100\% in the K-band.

\item Unresolved power-law and black-body functions contributions to the continuum are detected at the nucleus at the level of 20\,\% in the K-band. The first is attributed to the accretion disk emission and the latter to the emission from the dusty torus. We derive  a hot dust mass of $\sim 0.013\,$M$_{\odot}$, consistent with values observed for other Seyfert 2 galaxies.

\item The distribution of intermediate age stars shows a weak correlation with locations where we observe low stellar velocity dispersion values, supporting  that these low-$\sigma$ structures are originated in stellar populations that still preserve the cold kinematics of the gas from which they were formed.

\item By comparing the radial distribution of each stellar population component observed in Mrk~573 with those of other three galaxies studied using similar data (Mrk\,1066, Mrk\,1157 and NGC\,1068), we found that young stellar populations (contributing with $\ge$ 20\% of the K-band continuum) are observed for all galaxies within the inner 200--300 pc, while intermediate age stars dominates the near-IR K-band continuum at distances between 100 and 500~pc. Old stellar populations are dominant for distances smaller than 300~pc.

\item The stellar population extinction is larger at the nucleus and to the southwest of it, where we also observe higher gas extinction, as derived from the Pa$\beta$/Br$\gamma$ line ratio. The high extinction region to the southwest coincides also with a region of strong hot molecular gas emission.

\end{itemize}

\section*{Acknowledgements}

This work is based on observations obtained at the Gemini Observatory, 
which is operated by the Association of Universities for Research in Astronomy,
Inc., under a cooperative agreement with the 
NSF on behalf of the Gemini partnership: the National Science Foundation (United States), the Science and Technology 
Facilities Council (United Kingdom), the National Research Council (Canada), CONICYT (Chile), the Australian Research 
Council (Australia), Minist\'erio da Ci\^encia e Tecnologia (Brazil) and South-EastCYT (Argentina).  
This research has made use of the NASA/IPAC Extragalactic Database (NED) which is operated by the Jet
 Propulsion Laboratory, California Institute of  Technology, under contract with the National Aeronautics and Space Administration.
This work has been partially supported by the Brazilian institution CNPq.
{\it M.R.D.} thanks financial support from CNPq. {\it R.A.R.} acknowledges support from FAPERGS (project N0. 12/1209-6) and CNPq (project N0. 470090/2013-8). {\it R.R.} thanks to CNPq and FAPERGS for partially funding this work.

\label{lastpage}

\end{document}